\documentstyle[prd,aps,epsfig]{revtex}
\def\lsim{\mathrel{\rlap{\lower4pt\hbox{\hskip1pt$\sim$}}
    \raise1pt\hbox{$<$}}}               % less than or approx. symbol
\def\gsim{\mathrel{\rlap{\lower4pt\hbox{\hskip1pt$\sim$}}
    \raise1pt\hbox{$>$}}}               % greater than or approx. symbol

\begin{document}
\title{Quark-binding effects in inclusive decays of heavy mesons} 
\author{
Dmitri Melikhov$^{a}$\footnote{Alexander-von-Humboldt fellow;  
on leave from {\it Nuclear Physics Institute, Moscow State University, 
119899, Moscow, Russia.}}
and Silvano Simula$^{b}$}
\address{
$^a$ ITP, Universit\"at Heidelberg, Philosophenweg 16, D-69120, Heidelberg, Germany\\
$^b$ INFN, Sezione Roma III, Via della Vasca Navale, 84, I-00146 Roma, Italy}
\maketitle
\begin{abstract} 
We present a new approach to the analysis of quark-binding effects in 
inclusive decays of heavy mesons within the relativistic dispersion 
quark model. Various differential distributions, such as  
electron energy spectrum, $q^2$- and $M_X$-distributions, are calculated 
in terms of the $B$ meson soft wave function which also determines 
long-distance effects in exclusive transition form factors. 
Using the quark-model parameters and the $B$ meson wave
function previously determined from the description of the exclusive 
$b \to u$ transitions within the same dispersion approach, we provide numerical
results on various distributions in the inclusive $B \to X_c \ell \bar{\nu}_{\ell}$ decays. 
\end{abstract}
\section{Introduction}
Inclusive decays provide a promising possibility to determine 
the CKM matrix elements describing the mixing of $b$ quark, since a rigorous 
theoretical treatment of these decays, including nonperturbative effects, is possible. 
A consideration based on the Operator Product Expansion (OPE) 
and the Heavy Quark (HQ) expansion \cite{cgg}
allows one to connect the rate of the inclusive $B$ meson decay with
the rate of the $b$ quark decay. An important
consequence of the analysis based on the 
OPE is the appearance of the HQ binding effects 
in the integrated rates, both total and semileptonic (SL), 
of heavy meson decays 
only in the second order of the HQ expansion \cite{bsuv93}. These  
second order corrections are expressed in terms of the two hadronic 
parameters, $\lambda_1$ and $\lambda_2$. The latter are the mesonic matrix 
elements of the operators of dimension 5 which appear in the OPE of the 
product of the two weak currents. The differential distributions can be 
calculated as expansions in inverse powers of HQ mass $m_Q$ \cite{bsuv93,mw,fls}. 

Whereas presumably providing quite reliable results for the integrated SL decay rate, 
the OPE method encounters difficulties in calculating various 
differential distributions. For instance, before comparing the OPE-based results for the 
differential distributions with the true distributions a proper smearing 
over duality interval is necessary. 

There are several reasons which yield  
complications in the calculation of some differential distributions, 
arising mostly in the resonance region near zero recoil, namely:

\vspace{.25cm}
\noindent 1. the duality-violating $1/m_Q$ effects (i.e. the difference between the true distributions 
and the smeared OPE results) in the differential distributions 
near zero recoil which originate from the delay in opening different 
hadronic channels, as noticed in \cite{isgur}. Although these effects 
are cancelled in the integrated SL rate, they can considerably influence 
the kinematical distributions near the zero recoil point;

\vspace{.25cm}
\noindent 2. the convergence of the OPE series for the differential  
distributions persists only in the region where the quark 
produced in the SL decay is sufficiently fast. This means that the OPE 
cannot directly predict distributions in some kinematical regions, such as:  

\begin{itemize}
\item
the photon energy spectrum $d\Gamma/dE_\gamma$ in the radiative 
$B\to X_s\gamma$: the window in the photon 
energy between $m_Q/2$ and $M_Q/2$ turns out to be completely 
inaccessible within the OPE formalism \cite{fls}; 
\item
the lepton energy spectrum $d\Gamma/dE_{\ell}$ at large values of 
$E_{\ell}$ in semileptonic or rare leptonic decays; 
\item
the lepton $q^2$-distributions in SL $B \to X_c ~(X_u)$ and rare 
$B \to X_s$ decays at 
large $q^2$ near zero recoil; in this region one encounters both the  
quark-binding and duality-violating effects. 
\end{itemize}

Problems related to the quark-binding effects can be solved in principle
by performing proper resummation of the nonperturbative corrections 
which in practice however leads to the appearance 
of {\it a priori} unknown distribution functions \cite{neubert,bsuv94}. 

The inclusion of  
the quark-binding effects in heavy meson decays was 
first done in \cite{altarelli}, where an unknown distribution function of a 
heavy quark inside the heavy meson was introduced. Evidently, this 
distribution function is connected with the wave function of the heavy meson 
which also determines the exclusive transition form factors. To put this 
connection on a more solid basis, it is reasonable to consider the inclusive 
and exclusive processes within the same approach. 

We argue in this paper that the constituent quark model (QM) can be used 
an efficient tool for calculating differential distributions in inclusive 
decays of heavy mesons, covering also kinematical regions where 
OPE cannot provide a rigorous treatment. 
Namely, the constituent quark model 
allows one to take into account quark-binding effects in inclusive heavy meson 
decays in terms of the meson soft wave function. The latter describes the heavy 
meson properties both in exclusive and inclusive 
processes and thus allows one to consider on the same ground 
long-distance effects in various kinds of hadronic processes. 

Quark-model calculations of inclusive distributions are essentially based 
on the evaluation of the box diagram (see Fig. 1a later on) by introducing the heavy meson
wave function in one way or another. To illustrate the basic features 
of such an approach as well as its advantages and limitations it suffices to 
consider the case of a nonrelativistic (NR) potential model with scalar 
currents. Inclusion of relativistic effects can be then performed. 

Let us consider a weak transition induced by the scalar current $J=\bar c b$, where 
both $b$ and $c$ are heavy. To make the nonrelativistic treatment consistent we 
assume that 
\begin{eqnarray}
m_b, m_c\gg \delta m\equiv m_b-m_c \gg \Lambda,
\end{eqnarray}
where $\Lambda$ is the typical scale of the quark binding effects in the heavy meson.   
In the NR theory the general expression for the hadronic tensor 
\begin{eqnarray}
\label{w}
W(q_0,\vec q)&=&\frac{1}{\pi}{\rm Im}\int <B|T(J(x)J^+(0))|B>e^{-iqx}dx  
\end{eqnarray}
is reduced to the form 
\begin{eqnarray}
\label{wnr}
W(q_0,\vec q)=\frac{1}{\pi}{\rm Im}<B|G_{c \bar{d}}(M_B-q^0-i0,\vec q)|B>.  
\end{eqnarray}
Here $G_{c \bar{d}}(E,\vec q)=(\hat{H}_{c \bar{d}}(\vec q)-E)^{-1}$ 
is the full Green function corresponding to the the full Hamiltonian operator 
of the $c\bar d$ system with the total momentum $\vec q$ 
\begin{eqnarray}
\hat H_{c \bar{d}}(\vec q)=m_c+m_d+\frac{(\hat{\vec k}+\vec q)^2}{2m_c}+\frac{\hat{\vec k^2}}{2m_d}
+V_{c \bar{d}}(\hat r). 
\end{eqnarray}

Thus, the hadronic tensor is the average of the full $c\bar d$ Green function 
over the ground state of the full $b\bar d$ Hamiltonian 
\begin{eqnarray}
\label{eigen}
\hat{H}_{b \bar{d}}|B>=M_B|B>=(m_b+m_d+\epsilon_B)|B>.
\end{eqnarray}
In the rest frame of the $B$-meson one has  
\begin{eqnarray}
\hat H_{b \bar{d}}=m_b+m_d+\frac{\hat{\vec k^2}}{2m_b}+\frac{\hat{\vec k^2}}{2m_d}
+V_{b \bar{d}}(\hat r)
\equiv m_b+m_d+\hat h_{b \bar{d}}   
\end{eqnarray}
The following relation provides a basis for performing the 
OPE in the NR potential model \cite{raynal}  
\begin{eqnarray}
\label{basic-ope}
\hat H_{c \bar{d}}-(M_B-q^0)=
-(\delta m-\frac{\vec q^2}{2m_c}-q^0)+\left[\hat h_{b \bar{d}}-\epsilon_B\right]+
\frac{\hat{\vec k^2}+\hat V_1}{2}\left(\frac{1}{m_c}- \frac{1}{m_b}\right)
-\frac{\hat{\vec k}\vec q}{m_c}+O\left(\frac{\gamma^3\delta m}{m_c^3}\right)
\end{eqnarray}
where $\gamma \sim \Lambda_{QCD}$ and we have assumed the following expansion of the potential 
\begin{eqnarray}
\label{Vpot}
\hat V_{Q \bar{q}}=\hat V_0+\frac{\hat V_1}{2m_Q}+\frac{\hat V_2}{2m_Q^2}+...  
\end{eqnarray}

Starting with (\ref{basic-ope}) one constructs an OPE series using the 
amplitude of the 
free $b\to c$ quark 
transition as a zero-order approximation (hereafter referred to as the standard OPE). 
By virtue of the equations of motion, $\left(\hat h_{b \bar{d}}-\epsilon_B\right)|B>=0$ 
one observes the absence of $1/m_Q$ corrections 
to the leading order (LO) $b\to c$  amplitude,  
so that the corrections emerge only at the $1/m_Q^2$ order. 
Being completely reliable for the calculation of the integrated decay rate, this 
choice of the zero-order approximation turns out to be inconvenient however 
for calculating differential distributions. In particular, the distribution 
in the invariant mass of the produced hadronic system, $M_X$, becomes very 
singular and is represented via $\delta(M_X - m_c)$ and its derivatives, 
such that the $1/m_Q$ corrections are even more singular than the LO result. 
This is the price one pays for the choice of the zero-order term. 

It is clear that the free-quark transition amplitude is not the 
unique choice of the zero-order approximation, at least in quantum mechanics. 
For instance, another structure of the expansion 
can be obtained if the free $c\bar d$ Green function is used as the zero-order approximation.  

In the NR quantum mechanics the relation between the full and the free  
Green functions is well known and reads
\begin{eqnarray}
G^{-1}(E)=H-E, \quad G^{-1}_0(E)=H_0-E, \quad G^{-1}(E)-G^{-1}_0(E)=V, 
\end{eqnarray}
or, equivalently,  
\begin{eqnarray}
\label{exp2}
G(E)=G_0(E)-G_0(E)VG(E)=G_0(E)-G_0(E)VG_0(E)+G_0(E)VG_0(E)VG_0(E)+...
\end{eqnarray} 
For the heavy quark decay, in most of the kinematical $q^2$-region except for a 
vicinity of the zero recoil point, the Green function $G_0$ behaves as $1/m_Q$, and, since the matrix elements of the operator $V$ remain 
finite as $m_Q \to \infty$, the series 
(\ref{exp2}) is an expansion in powers of $1/m_Q$. 
Notice that in the NR potential model the expansion (\ref{exp2}) is fully equivalent 
to the OPE series obtained from (\ref{basic-ope}). 
Inserting the expansion (\ref{exp2}) into the expression for the hadronic tensor $W$ given by 
eq. (\ref{wnr}), we 
come to the 
series shown in Fig 1. 
\begin{figure}[htb]
\vspace{0.25cm}
\begin{center}
\epsfig{file=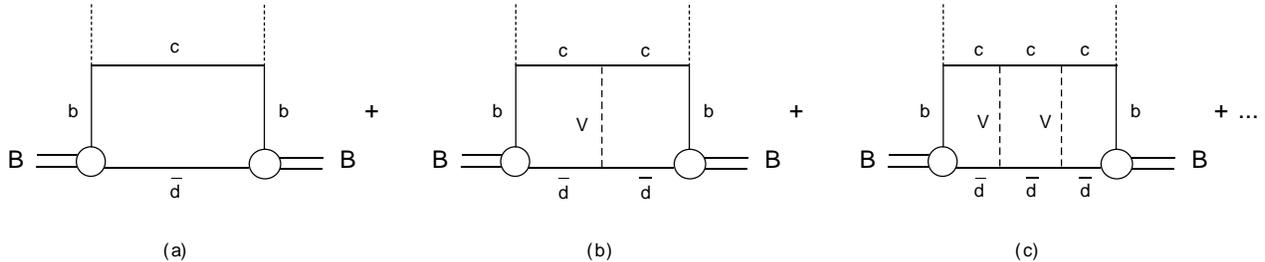,width=16.75cm}
\vspace{0.25cm}
\caption{ Expansion of the hadronic tensor in the quark model: 
(a) - the box diagram which provides the LO free-quark contribution, 
(b,c) - diagrams contributing in subleading $1/m_Q$ orders and containing the final state interaction $V \equiv V_{c \bar{d}}$. The short-dashed lines represent W bosons. 
\label{fig:fig1}}
\end{center}
\end{figure}
The LO term is the box diagram of Fig. 1a  
with the free $c$ and $\bar{d}$ quarks in the 
intermediate state. The corresponding analytical expression reads 
\begin{eqnarray}
W^{QM}=\frac{1}{\pi}{\rm Im}<B|G^0_{c \bar{d}}(M_B-q^0,\vec q)|B>.   
\end{eqnarray}
This is the quantity usually taken into account in QM calculations. 
It is easy to see, that the SL decay rate calculation based on the box diagram of Fig. 1 only,  
reproduces the free quark SL decay rate in the HQ limit, but should contain also $1/m_Q$ 
corrections (see also the general structure of the QM results of Ref. \cite{termarti}). Namely, the next-to-leading order (NLO) term in the expansion (\ref{exp2}) is the diagram 
with a single insertion of the potential between the free $c$ and $\bar{d}$ quarks 
(the diagram of Fig. 1b). 
It has the order $1/m_Q$ and precisely cancels the $1/m_Q$ contribution of the QM box diagram, 
yielding in this way the absence of the $1/m_Q$ correction in the difference between the decay 
rates of bound and free heavy quarks.   

Thus, the QM box-diagram calculation is just the first--order 
term in an alternative expansion of the full Green function: unlike the 
standard OPE series which starts from a single $c$-quark in the intermediate 
state, 
the QM starts from the free $c\bar d$ pair which is the eigenstate 
of the Hamiltonian
\begin{eqnarray}
H^0_{c \bar{d}}(\vec q)=m_c+m_d+\frac{(\vec k+\vec q)^2}{2m_c}+\frac{\vec k^2}{2m_d}. 
\end{eqnarray} 
Hereafter we refer to the expansion of the decay rate based on the expansion 
(\ref{exp2}) of the Green function as the QM expansion. 

Summing up, the quark model provides an alternative $1/m_Q$ expansion with the following properties: 
\begin{itemize}

\item[1.] the box diagram of Fig. 1a provides the LO $1/m_Q$ term and reproduces the 
free-quark decay in the limit $m_Q\to\infty$. All other terms contribute only 
in subleading $1/m_Q$ orders;

\item[2.] the differential distributions in any $1/m_Q$ order are convergent 
for almost all allowed $q^2$, except for the region close to zero recoil; 

\item[3.]  before comparing the calculated differential distributions 
based on the expansion (\ref{exp2}) with the true distributions 
in the resonance region a proper smearing over some duality interval is required;

\item[4.] the $1/m_Q$ correction to the LO term is nonvanishing.

\end{itemize}

Clearly, the properties 1-3 are completely equivalent to the standard OPE,  
while the property 4 makes the QM expansion much less convenient than the 
standard OPE, at least for the calculation of the SL decay rate. However:
\begin{itemize} 

\item[5.] the expansion (\ref{exp2}) turns out to be more suitable for the calculation of 
the differential distributions, e.g. for the calculation of $d\Gamma/dM_X$: 
in this case the LO result (the box diagram of Fig. 1a) 
is well-defined in the whole kinematical region as well as the higher 
order corrections to it. Thus already the box diagram is appropriate for comparison 
with the experimental $d\Gamma/dM_X$ at all $M_X$ apart from the 
resonance region. 
Beyond the resonance region no additional smearing of the calculated 
$d\Gamma/dM_X$ is required.

\end{itemize}

In full QCD the situation is of course much more complicated and hadrons are 
coherent states of infinite number of quarks and gluons. Nevertheless, many 
applications of the constituent quark model have proved the treatment of mesons 
as bound states of two constituent quarks to provide a reasonable description 
of their properties. From this viewpoint the arguments given above remain valid. 
Namely, the box diagram represents the main contribution to the hadronic tensor
which reproduces the free-quark decay in the infinite quark mass limit. 
However hadronic tensor calculated from the box diagram contains the $1/m_Q$ term 
compared with the free-quark decay tensor. This linear $1/m_Q$ term is known to be cancelled 
by the $1/m_Q$ order contributions of higher order diagram and to be absent in the full
expression. In practice, however the 
$1/m_Q$ term of the box diagram is not so dangerous: namely,   
the hadronic tensor calculated from the box diagram, as well as all corrections given by the 
other graphs, are regular in the whole kinematical region. Thus {\it the box diagram 
should provide a reasonable description already appropriate for comparison 
with experiment}. Moreover, the box-diagram result 
can be further improved by effectively taking 
into account the higher order term which kills the $1/m_Q$ 
correction contained in the box diagram. We follow this strategy in our analysis and perform a
relativistic treatment of quark-binding effects within a constituent quark picture. 

Our consideration of the quark binding effects in inclusive SL decays is 
based on the relativistic dispersion formulation of the quark model previously developed for 
the description of meson transition form factors \cite{m}. 
Within this approach,  
the inclusive decay rates as well as the exclusive hadron transition form 
factors are given by double spectral representations in terms of the 
soft meson wave functions. The double spectral densities of these  
spectral representations are obtained from the corresponding Feynman graphs.  
The subtraction terms in spectral representations for exclusive 
transition form factors are fixed by requiring the structure of the HQ expansion in the QM to match the structure of the HQ expansion in QCD. 
In this paper we proceed along the same lines in inclusive processes. 

Our main results are the following:
\begin{itemize}

\item we construct the double spectral representation of the hadronic tensor 
within the constituent quark model starting with $q^2<0$. The hadronic tensor 
is represented in terms of the soft wave function of the $B$ meson and the 
double spectral density of the box diagram. The hadronic tensor at $q^2>0$ 
is obtained by the analytical continuation. 
Then the $1/m_Q$ expansion of the spectral representation of the 
decay rate is performed and the LO term is shown to reproduce the free 
quark decay rate. The subtraction is defined in such a way  
that the $1/m_Q$ correction to the SL decay rate is absent. 
This corresponds to effectively taking into account other terms  
beyond the box-diagram approximation which contribute in subleading $1/m_Q$ orders. 
Moreover an approximate account of the $1/m_Q^2$ effects of the whole series 
within the
box-diagram expression is possible. This is done by introducing a phenomenological cut in the double
spectral representation of the box-diagram which affects only the differential
distributions at large $q^2$. This cut brings the size of the $1/m_Q^2$ corrections in the 
$\Gamma(B \to X_c \ell \bar{\nu}_{\ell})$ in full agreement with the OPE result and keeps the LO and $1/m_Q$ 
correction unchanged. The cut yields differential distributions which are finite also 
in the endpoint $q^2$-region where the HQ expansion series is not properly convergent;

\item various differential distributions are calculated in terms of the $B$-meson soft 
wave function. These distributions are regular in the whole kinematically accessible region 
and, apart from the 
resonance region (where the exact distributions are dominated by single resonances and 
a proper smearing over the duality intervals is necessary), can be directly compared with the observable values. 
The main effect of quark-binding upon these distributions 
is determined unambiguously through the soft wave function, while the $1/m_Q^2$ corrections 
depend on the particular details of an approximate account of the higher-order terms in the 
series (\ref{exp2}).  
However, in practice these 
details are not essential due to the following two reasons: 
first they are numerically small, and 
second, the size of the $1/m_Q^2$ corrections in the integrated SL rate is close to the OPE result. 
So we expect that the size of the $1/m_Q^2$ corrections is 
reasonably reproduced also in other quantities;

\item we perform numerical estimates of various differential 
distributions in inclusive decays 
in terms of the $B$-meson wave function known from the 
description of the exclusive processes \cite{mns,bm}. 

\end{itemize}

The paper is organized as follows: in the next section we present necessary formulas
for the free-quark decay and also the OPE prediction for the total integrated rate up to $1/m_Q^2$ corrections. In Section 3 we construct the dispersion representation 
for the box diagram at $q^2<0$ and discuss its analytical continuation to
$q^2>0$. Section 4 performs the $1/m_Q$ expansion of the hadronic tensor in the quark model, 
and Section 5 presents numerical results for 
differential distributions. Finally, a brief summary is given in the Conclusion. 

\section{Free quark decay and OPE}

Let the effective Hamiltonian governing the quark transition $Q\to Q'$ 
with the emission of the particle $\phi$ have the following structure  
\begin{eqnarray}
\label{heff}
H_{eff}(x)=\bar Q'(x)\hat\Gamma Q(x)\phi(x)  
\end{eqnarray}
where $\hat\Gamma$ denotes a relevant combination of the Dirac 
matrices. Following notations of ref \cite{m} for the exclusive form factors, 
we denote the parent heavy quark $Q$ also as $Q_2$ and the 
daughter quark $Q'$ as $Q_1$. 

A tree-level rate of the free-quark decay initiated by this effective
Hamiltonian averaged over the
polarizations of the initial quark $Q_2$ and summed over polarizations of
the final quark $Q_1$ has the form
\begin{eqnarray}
\label{gamma0}
{d\Gamma_0 \over dq^2}&=&\frac{(2\pi)^4}{2m_2}\int |T_0|^2 
\frac{dk_1 dk_\phi}{(2\pi)^6} \delta(k_1^2-m_1^2)
\delta(k_\phi^2-q^2) 
\delta(k_2-k_1-k_\phi)  \nonumber \\
&=&\frac{(2\pi)^4}{2m_2(2\pi)^6}\frac{\pi\lambda^{1/2}(m_2^2,m_1^2,q^2)}{2m_2^2}
|T_0|^2, 
\end{eqnarray}
where $q^2$ is the mass squared of the particle $\phi$ and 
\begin{eqnarray}
\label{free}
|T_0|^2=\frac{1}{2}\sum_\sigma \bar u_\sigma (k_2)\hat \Gamma 
(m_1+\hat k_1)\hat\Gamma u_\sigma(k_2)=
\frac{1}{2}{\rm Sp} \left( (m_2+\hat k_2)\hat\Gamma (m_1+\hat k_1)
\hat\Gamma\right). 
\end{eqnarray}
Hereafter we use the notation $\lambda(x,y,z)\equiv (x+y-z)^2-4xy$.

These formulas can be readily applied to the particular cases of  
radiative and SL decays. In the latter case one needs to multiply $d\Gamma_0/dq^2$ by the 
the leptonic tensor $L(q^2)$ to obtain 
the full differential distribution $d\Gamma_0^{SL}/dq^2=d\Gamma_0/dq^2 \cdot L(q^2)$. Hereafter the inclusion of the 
leptonic tensor in the definition of $d\Gamma_0/dq^2$ is understood and we drop the superscript SL. 

In the case of the SL $b \to c \ell \bar{\nu}_{\ell}$ transition the 
free-quark differential decay rate reads explicitly as 
(cf., e.g., \cite{bsuv93,mw})
\begin{equation}
\label{freeq}
{d\Gamma_0 \over dq^2} = {G_F^2 |V_{bc}|^2 \over 96 \pi^3} \frac{1}{m_b^3} 
 \lambda^{1/2}(m_b^2,m_c^2,q^2) C(m_b^2,m_c^2,q^2),
\end{equation}
where $G_F$ is the universal Fermi constant, $V_{bc}$ is the CKM matrix element, 
and 
\begin{equation}
\label{c}
C(m_b^2,m_c^2,q^2)=(m_b^2-m_c^2)^2+q^2(m_b^2+m_c^2)-2q^4. 
\end{equation}
The integrated SL rate is then given by
\begin{equation}
 \label{freeq_total}
 \Gamma_0 = \int_0^{(m_b-m_c)^2} dq^2 {d\Gamma_0 \over dq^2} = {G_F^2 
 |V_{bc}|^2 \over 192 \pi^3} \cdot m_b^5 \cdot I_0(r)
\end{equation}
with $I_0(r) = 1 - 8 r + 8 r^3 - r^4 - 12 r^2 {\rm ln}(r)$, $r \equiv (m_c / m_b)^2$. 

The OPE predicts that in the decay of the heavy meson, the $1/m_b$ corrections to 
the free quark rate (\ref{freeq_total}) are absent, and 
up to $1/m_b^2$ terms the integrated SL rate is given by (cf., e.g., \cite{bsuv93,mw})
\begin{equation}
 \label{OPE_total}
 \Gamma = \Gamma_0 \cdot \left[ 1 + {\lambda_1 + 3 \lambda_2 \over 2 m_b^2} 
 - 6 {\lambda_2 \over m_b^2} {(1 - r)^4 \over I_0(r)} \right].
\end{equation}
Here $\lambda_1$ and $\lambda_2$ are the hadronic matrix elements of the operators of 
dimension 5 appearing in the OPE of the product of the two weak currents. 
The value $\lambda_2 = 0.12 ~ GeV^2$ is well known from the $B-B^*$ mass splitting,
whereas the knowledge of $\lambda_1$ is loose and present estimates range from $-0.6 ~ GeV^2$ to $0$. Note that in the NR quark potential model one has $\lambda_1 = -<\vec{k}^2>$, where $\vec{k}$ is the relative momentum of the constituent $Q \bar{q}$ pair (cf, e.g., \cite{sim}). Typically, the NR quark model estimates of $\lambda_1$ range from $-0.6$ to $-0.4 ~ GeV^2$.

\section{Inclusive meson decay in the quark model}

We now proceed to the calculation of the inclusive rate for the decay of a pseudoscalar
meson with mass $M_1$ containing a heavy quark $Q_2$, which we
will refer to as $P_{Q_2}$, induced by the quark transition (\ref{heff}).

\begin{figure}[htb]
\vspace{0.25cm}
\begin{center}
\epsfig{file=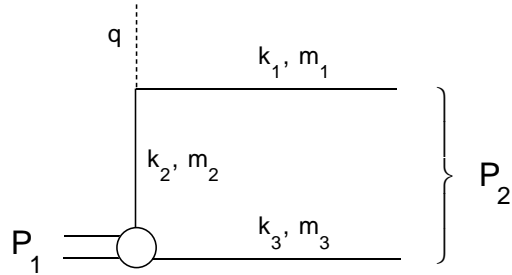,width=7cm}
\vspace{0.25cm}
\caption{Momentum notations in the box diagram. 
\label{fig:notation}}
\end{center}
\end{figure}

We start with the box diagram of Fig. 1a.
Our notations shown in Fig \ref{fig:notation} follow those of ref \cite{m} where transition form
factors within the similar dispersion approach have been considered. 
The decay rate corresponding to this diagram can be written in the form  
\begin{equation}
{d\Gamma(P_{Q_2}\to X_{Q_1}\phi) \over dq^2}=
\frac{(2\pi)^4}{2M_1}\int \frac{dk_1 dk_3 dk_\phi}{(2\pi)^9}|T|^2
\delta(k_1^2-m_1^2)
\delta(k_3^2-m_3^2)
\delta(k_\phi^2-q^2)\delta(p_1-k_2-k_3-k_\phi), 
\end{equation}
where $T$ is the amplitude of the meson decay
$P_{Q_2}\to Q_1\bar q_3\phi$. 

In the dispersion approach the decay rate can be written in the 
form of the following spectral representation: 
\begin{equation}
\label{g1}
{d\Gamma(P_{Q_2}\to X_{Q_1}\phi) \over dq^2}=
\frac{(2\pi)^4}{2M_1}\int \frac{ds_1\;ds_2}{(s_1-M_1^2)^2}
\frac{\pi\lambda^{1/2}(s_1,s_2,q^2)}{2s_1}\tilde A(s_1,s_2,q^2)
\end{equation}
where the spectral density $\tilde A(s_1,s_2,q^2)$ 
is connected with the double discontinuity of the 
box diagram shown in Fig. 1. 

The amplitude $A$ of Fig. 1a depends on six Lorentz scalar
variables $p_1^2,p'^2_1,p^2_2,q^2,q'^2,\kappa^2$, where the momenta
satisfy the following relations 
$p_1-p'_1=q-q'=\kappa$,  and $p_1-q=p'_1-q'=p_2$. The amplitude can
be written as double spectral representation in $p_1^2$ and $p'^2_1$ 
\begin{equation}
\label{drfora}
A(p_1^2,p'^2_1,p^2_2,q^2,q'^2,\kappa^2)=\int \frac{ds_1}{s_1-p_1^2}
\frac{ds'_1}{s'_1-p'^2_1}\tilde A(s_1,s'_1,p^2_2,q^2,q'^2,\kappa^2). 
\end{equation}
In this equation $\tilde A$ is the full spectral density of the amplitude 
and includes properly defined subtraction terms. The spectral density $\tilde A$ is calculated 
from the double discontinuity $\tilde A_D$,  
\begin{equation}
\tilde A_D=
\frac{1}{(2\pi i)^2}{\rm disc}_{s_1}{\rm disc}_{s'_1}A(s_1,s'_1,p^2_2,q^2,q'^2,\kappa^2), 
\end{equation}
through a relevant subtraction procedure. 

\subsection{The spacelike region}

The double discontinuity $\tilde A_D$ of the box diagram of Fig. 
1a can be calculated at $q^2\le 0$ and $q'^2\le 0$ by placing all
intermediate quarks on their mass shells but keeping the initial and 
final mesons virtual and having the masses squared $s_1$ and $s_1'$,
respectively. 
To obtain the corresponding expression at positive values of
$q^2=q'^2$ which is necessary e.g. for SL decays, we shall perform the 
analytical continuation in $q^2$. 
At $q^2, ~ q'^2\le 0$ the procedure explained in detail in \cite{m} yields the
expression 
\begin{eqnarray}
\tilde A_D(s_1,s'_1,s_2,q^2,q'^2,\kappa^2) &=& \frac{1}{(2\pi)^9} 
\int dk_1 dk_2 dk'_2 dk_3 \delta(k_1^2-m_1^2) 
\delta(k_2^2-m_2^2)\delta(k'^2_2-m_2^2)\delta(k_3^2-m_3^2)  \nonumber\\
&&\times 
\delta(\tilde p_1-k_2-k_3)\delta(\tilde p'_1-k'_2-k_3)
\delta(\tilde p_2-k_1-k_3) \nonumber\\
&&\times 
(-1){\rm Sp}\left(i\gamma_5\,G(s_1)\,(m_3-\hat k_3)
i\gamma_5\,G(s'_1)\,
(m_2+\hat k_2)\hat\Gamma(m_1+\hat k_1)\hat\Gamma(m_2+\hat
k_2)\right),    
\end{eqnarray} 
where the momenta satisfy the following relations 
$\tilde p'_1=\tilde p_1+\kappa$, 
$\tilde q'=\tilde q+\kappa$,  
$\tilde p_2=\tilde p_1-\tilde q$, 
$\tilde p_1^2=s_1$, $\tilde p'^2_1=s'_1$, $\tilde q^2=q^2$,  
$\tilde q'^2=q'^2$, $\tilde p^2_2=s_2$. 
Notice that the quark structure of the (virtual) pseudoscalar meson transition
into two real quarks is described by the vertex 
$G(s_1) \bar Q_2(k_2)i\gamma_5 q_3(k_3)\delta(\tilde p_1-k_2-k_3)$,
where $\tilde p_1^2=s_1, k_2^2=m_2^2, k_3^2=m_3^2$. 

The spectral density $\tilde A(s_1,s_2,q^2)$ in (\ref{g1}) is 
connected with the forward double spectral density of the box diagram 
$\tilde A(s_1,s_1',s_2,q^2,q'^2,\kappa^2)$ as follows
\begin{eqnarray}
\lim_{\kappa^2\to 0}\;\tilde A(s_1,s_1',s_2,q^2,q^2,\kappa^2)=
\delta(s_1-s_1')\tilde A(s_1,s_2,q^2).
\end{eqnarray}
The form of the subtraction procedure in the spectral representation 
(\ref{drfora}) cannot be determined within the dispersion approach and should
be fixed from some other arguments. 
We can determine the subtraction term 
by requiring the absence of the $1/m_Q$ corrections in the ratio of the bound 
and free quark decay rates. As we have discussed, this corresponds to 
taking into account the $1/m_Q$ terms of other diagrams which are known
to cancel the $1/m_Q$ term in the bix diagram. Our subtraction prescription explicitly reads 
\begin{eqnarray}  
\label{prescr}
\tilde A(s_1,s_2,q^2)=
\frac{M_1}{\sqrt{s_1}}\tilde A_D(s_1,s_2,q^2)  
\end{eqnarray}
where $\tilde A_D$ is connected with the double discontinuity of the 
forward amplitude 
\begin{eqnarray}
\lim_{\kappa^2\to 0}\;\tilde A_D(s_1,s_1',s_2,q^2,q^2,\kappa^2)=
\delta(s_1-s_1')\tilde A_D(s_1,s_2,q^2).
\end{eqnarray}
Explicit calculations give for $\tilde A_D$ the following expression
\begin{eqnarray}
\label{ad}
\bar A_D(s_1,s_2,q^2)
=\frac{\pi\theta(...)}{2\lambda^{1/2}(m_1^2,m_2^2,q^2)}
(-{\rm Sp}). 
\end{eqnarray}
Notice that the argument of the $\theta$ function in eq (\ref{ad}) is just the 
same as for the spectral density of the triangle graph describing the form 
factor in the dispersion approach which can be read off from \cite{m}. 
Since all quarks are on their mass shell, the trace can be rewritten
in the following form 
\begin{eqnarray}
-{\rm Sp}=2|T_0|^2(s_1-(m_2-m_3)^2), 
\end{eqnarray}
where $|T_0|^2$ is just the square of the free-quark amplitude of eq 
(\ref{free}). 
Finally, isolating the free-quark decay amplitude and solving the
kinematical $\theta$ function we come to the following dispersion 
representation for the differential inclusive decay rate at $q^2\le 0$
\begin{eqnarray}
\label{inclrate}
{d\Gamma \over dq^2} = K_0(q^2)
\int ds_1 \frac{G^2(s_1)}{(s_1-M_1^2)^2}\frac{s_1-(m_2-m_3)^2}{8\pi^2s_1}
\frac{m_2}{\sqrt{s_1}}
\frac{m_2^2}{\lambda^{1/2}(m_1^2,m_2^2,q^2)}
\int_{s_2^{-}(s_1,q^2)}^{s_2^{+}(s_1,q^2)}ds_2\lambda^{1/2}(s_1,s_2,q^2), 
\end{eqnarray}
where
\begin{eqnarray}
K_0(q^2)\equiv\frac{1}{\lambda^{1/2}(m_2^2,m_1^2,q^2)}\frac{d\Gamma_0}{dq^2}=
\frac{G_F^2|V_{21}|^2}{96\pi^3}\frac{C(m_2^2,m_1^2,q^2)}{m_2^3}. 
\end{eqnarray}

The limits $s_2^{\pm}$ are obtained by setting $\eta=\pm 1$ in the equation 
\begin{eqnarray}
\label{limits}
s_2(s_1,q^2)&=&(m_1+m_3)^2+
\frac{m_1}{m_2}\left(s_1-(m_2+m_3)^2    \right)
+\frac{m_1}{m_2}(\omega -1)(s_1-m_2^2-m_3^2) \nonumber \\
&+&\eta\frac{m_1}{m_2}\lambda^{1/2}(s_1,m_2^2,m_3^2)\sqrt{\omega^2-1},
\end{eqnarray}
where the quark recoil $\omega$ is defined as follows
\begin{eqnarray}
\label{omega}
q^2=(m_2-m_1)^2-2m_1m_2(\omega-1).
\end{eqnarray}
Note that in eq (\ref{inclrate}) the free-quark differential rate $d\Gamma_0 / dq^2$ 
factorizes out, so that the differential rate for a bound quark is a product of the free-quark 
differential rate and a bound state factor, as already noted in \cite{termarti}.
Hereafter we use the notation $\varphi(s)=G(s)/(s-M_1^2)$. 
The normalization condition of the
soft wave function $\varphi(s)$ obtained from the 
elastic vector form factor of the heavy meson at $q^2=0$ reads
\cite{m} 
\begin{eqnarray}
\label{norma}
\int ds_1 \varphi^2(s_1)
\left[s_1-(m_2-m_3)^2\right]\frac{\lambda^{1/2}(s_1,m^2_2,m_3^2)}{8\pi^2s_1}=1.
\end{eqnarray}
It is convenient to rearrange eq (\ref{inclrate}) by isolating
under the integral the structure similar to the structure of the
normalization condition (\ref{norma})
\begin{eqnarray}
\label{inclrate2}
{d\Gamma \over dq^2} & = & K_0(q^2)
\int ds_1 \varphi^2(s_1)
\left[s_1-(m_2-m_3)^2\right]\frac{\lambda^{1/2}(s_1,m^2_2,m_3^2)}{8\pi^2s_1}
\rho(s_1,q^2),\nonumber  \\
\rho(s_1,q^2) & = & \frac{m_2}{\sqrt{s_1}}
\frac{m_2^2}{\lambda^{1/2}(s_1,m^2_2,m_3^2)\lambda^{1/2}(m^2_2,m_1^2,q^2)}
\int_{s_2^-}^{s_2^+} ds_2\lambda^{1/2}(s_1,s_2,q^2).
\end{eqnarray}
As we shall see later, $\rho(s_1,q^2)\sim \lambda^{1/2}(m^2_2,m_1^2,q^2)$ 
in the HQ limit, so that thanks to the normalization (\ref{norma}) of the soft wave function one gets $d\Gamma / dq^2 \to d\Gamma_0 / dq^2$ as $m_Q \to \infty$. 

\subsection{The timelike region and the anomalous contribution}

To obtain the spectral representation at $q^2>0$ we perform the
analytical continuation in $q^2$. This procedure 
is done along the same lines as in the case of the transition form factor 
which has been discussed in
detail in \cite{m}. 
As a result of this procedure 
in addition to the normal part which is just the expression
(\ref{inclrate2}) taken at $q^2>0$, 
an anomalous part emerges due to the non-Landau type 
singularities of the Feynman graph. Thus, in the region $q^2\le (m_2-m_1)^2$ 
the representation for $\rho(s_1,q^2)$ takes the form 
\begin{eqnarray}
\rho(s_1,q^2)&=&\frac{m_2}{\sqrt{s_1}}
\frac{m_2^2}{\lambda^{1/2}(s_1,m^2_2,m_3^2)\lambda^{1/2}(m^2_2,m_1^2,q^2)}\nonumber 
\\
&&\times
\left[{
\int_{s_2^-(s_1,q^2)}^{s_2^+(s_1,q^2)} ds_2\lambda^{1/2}(s_1,s_2,q^2)
+2\theta(q^2)\theta(s_1>s_1^0)\int_{s_2^+(s_1,q^2)}^{s_2^L(s_1,q^2)} ds_2\lambda^{1/2}(s_1,s_2,q^2)
}\right], 
\end{eqnarray}
where
\begin{eqnarray}
s_2^L(s_1,q^2)&=&(\sqrt{s_1}-\sqrt{q^2})^2, \nonumber \\
\sqrt{s_1^0(q^2)}&=&\frac{q^2+m_2^2-m_1^2}{2\sqrt{q^2}}+\sqrt{
\left({\frac{q^2+m_2^2-m_1^2}{2\sqrt{q^2}}}\right)^2+m_3^2-m_2^2}.
\end{eqnarray}

The $q^2$-behavior of the anomalous
term is determined by the lower limit of the $s_1$ integration, $s^0_1(q^2)$. 
Namely, its contribution to the SL rate reads 
\begin{eqnarray}
\frac{1}{\Gamma}\frac{d\Gamma^{anom}}{d\omega}\simeq 
\frac{\Lambda^3}{m_Q^3\sqrt{\omega-1}}R^{anom}(\omega),
\end{eqnarray}
where 
\begin{eqnarray}
R^{anom}(\omega)=\int\limits_{\frac{m_Q^2}{\Lambda^2}(\omega-1)}^{\infty}
d\vec k^2 \Psi^2_B(\vec k^2).
\end{eqnarray}
Here $|\vec k|= \lambda^{1/2}(s_1,m_2^2,m_3^2) / 2\sqrt{s_1}$ 
is the relative momentum of the quarks inside the $B$-meson, and the 
wave function $\Psi_B(\vec k^2)$ can be expressed through $\varphi(s)$ 
\cite{m}. In terms of  
$\Psi_B(\vec k^2)$ the normalization condition (\ref{norma}) takes the form 
$\int \Psi_B^2(\vec k^2)d\vec k^2=1$ such that $R^{anom}(1)=1$.
Since the soft wave function is steeply falling beyond the 
confinement region where $\vec k^2\lsim\Lambda^2$, the anomalous contribution 
becomes inessential already at $\omega-1\simeq \Lambda/m_b$. 
Only in the endpoint region $\omega-1\lsim \frac{\Lambda^2}{m_Q^2}$ 
the anomalous contribution to the differential distribution becomes strong 
and diverges like 
$\frac{1}{\Gamma}\frac{d\Gamma^{anom}}{d\omega}\simeq \frac{\Lambda^3}{m_Q^3}
\frac{1}{\sqrt{\omega-1}}$. 

The contribution of the anomalous term to the integrated SL rate is  
$\Gamma^{anom}/\Gamma\simeq \Lambda^2/m_Q^2$ and comes 
from the endpoint region, whereas the rest of the phase space 
$0<q^2\lsim\Lambda/m_Q$ provides only the relative $\Lambda^3/m_Q^3$ 
anomalous contribution to the SL decay rate. 

Therefore, the anomalous contribution is negligible at all $\omega$ 
except for the endpoint region $\omega-1\lsim \frac{\Lambda^2}{m_Q^2}$, which 
is in fact a very narrow region near zero recoil. As we have discussed, the HQ expansion for the differential distributions 
is anyway ill-defined in this kinematical region. Contributions of 
the same order of magnitude come also from other terms in the expansion 
(\ref{exp2}), and keeping this anomalous contribution is beyond  
the accuracy of our considerations. 
Thus we shall systematically omit the anomalous contribution in numerical calculations. 

\section{The heavy quark expansion of the inclusive decay rate in the quark model}

In this section we perform the HQ expansion of the meson inclusive decay
rate. We show that:
\begin{itemize}

\item[a.] in the LO the heavy meson inclusive decay rate is equal to the free quark decay rate;

\item[b.] our subtraction prescription  (\ref{prescr}) leads to the 
differential distribution $d\Gamma/dq^2$ given by eq (\ref{inclrate2}) which satisfies the relation 
$(d\Gamma(B\to X_c \ell \bar{\nu}_{\ell})/dq^2)/(d\Gamma(b\to c \ell \bar{\nu}_{\ell})/dq^2)=
1+O(1/m_Q^2)$ in most of the $q^2$ region except for 
a close vicinity of zero recoil point. This property guarantees the absence of the $1/m_Q$ 
in the ratio of the integrated rates, i.e. 
$\Gamma(B\to X_c \ell \bar{\nu}_{\ell})/\Gamma(b\to c \ell \bar{\nu}_{\ell})=1+O(1/m_Q^2)$;

\item[c.] the size of the $1/m_Q^2$ corrections can be tuned such that they become numerically 
close to the OPE prediction. This is done by introducing the cut in the spectral 
representation of the decay rate of the $B$ meson. This cut affects only the 
differential distribution $d\Gamma(B \to X_c \ell \bar{\nu}_{\ell})/dq^2$ at large $q^2$ 
near zero recoil, i.e. in the region 
$\omega\le 1+O(1/m_Q)$. 
 
\end{itemize}

The most important feature of the whole approach is that already the zero order 
expression provides a realistic $M_X$-distribution. Modifications b) and c) while affecting the 
total rate and the $q^2$-distributions at large $q^2$, only moderately affect 
the $M_X$-distribution, so that the latter is mostly determined only by the soft 
$B$-meson wave function.

\subsection{Soft wave function and normalization condition} 
First, we need to specify the properties of the soft meson  wave
function. A basic property of the soft wave function
$\varphi(s,m_Q,m_{\bar q},\Lambda)$ is its strong peaking in
terms of the relative quark momentum in the region of the order of
$\Lambda \simeq \Lambda_{QCD}$. 

For elaborating the $1/m_Q$ expansion, it is convenient to formulate
such peaking in terms of the variable $z$ such that $s=(m_Q+m_3+z)^2$ 
(hereafter we denote the mass of the light spectator quark as $m_3$).
In the heavy meson, the variable $z$ is related to the relative quark
momentum as follows 
\begin{equation}
\label{z2k}
{\vec k}^2=z(z+2m_3)+O(1/m_Q).
\end{equation}
Hence, a localization of the soft wave function in terms of $z$ means that the 
wave function is nonzero as $z\le \Lambda$. 
In the heavy meson case we imply that $m_Q \gg m_{3}\simeq z\simeq
\Lambda$.

The normalization condition (\ref{norma}) is  
a consequence of the vector current conservation in the full theory
and it provides an (infinite) chain of relations at different $1/m_Q$ orders. 
Namely, expanding the soft wave function in $1/m_Q$ as follows 
\begin{equation}
\label{phi}
\varphi(s,m_Q,m_{3},\Lambda)=\frac\pi{\sqrt{m_Q}}\phi_0(z,m_{3},\Lambda)
\left[{1+\frac {m_3}{4m_Q}
\chi_1(z,m_{3},\Lambda)+O(1/m_Q^2)}\right],
\end{equation}
we come to the normalization condition in the form 
\begin{equation}
\int dz \phi^2_0(z)\sqrt{z}(z+2m_3)^{3/2}\left[1+\frac{m_3}{2m_Q}\chi_1(z)-\frac{m_3}{2m_Q}+\ldots\right]=1. 
\end{equation}
This exact relation is equivalent to an infinite chain of equations in different $1/m_Q$ orders. 
Lowest order relations take the form  
\begin{eqnarray}
\label{normaphi}
&\int &dz\;\phi^2_0(z)\sqrt{z}(z+2m_3)^{3/2}=1, \\
&\int & dz\;\phi^2_0(z)\sqrt{z}(z+2m_3)^{3/2}\chi_1(z)=1, etc. 
\end{eqnarray}
In particular, the Isgur-Wise function is given by the following
expression through $\phi_0$ 
\begin{eqnarray} 
\label{xi} 
\xi(\omega )=\int 
dz_1 \phi_0(z_1)\sqrt{z_1(z_1+2m_3)}\int\limits_{-1}^{1}\frac{d\eta}2
\phi_0(z_2)\left({m_3+\frac{2m_3+z_1+z_2}{1+\omega }}\right).  
\end{eqnarray} 
The expression for $z_2$ through $z_1$ and $\eta$ ($-1 \leq \eta \leq 1$) is obtained by
expanding (\ref{limits}) in $1/m_Q$   
\begin{eqnarray}
\label{limitsz}
z_2&=&z_1+(z_1+m_3)(\omega-1) + \eta\sqrt{z_1(z_1+2m_3)}\sqrt{\omega^2-1}+O(1/m_Q), 
\end{eqnarray}
and for the calculation of the IW function only the LO part of this relation should 
be used. 

\subsection{The HQ expansion}

First let us consider the HQ expansion of the unsubtracted quantity $\rho_D$. 
The normal part of $\rho_D(s_1,q^2)$ reads  
\begin{eqnarray}
\label{r2}
\rho_D(s_1,q^2)=\frac{m_2}{M_1}
\int_{-1}^{1}\frac{d\eta}{2}\lambda^{1/2}(s_1,s_2,q^2).
\end{eqnarray}
This representation is a convenient starting point for performing the HQ 
expansion. Notice that although the integration in $\eta$ can be
easily performed, it is more convenient to work out the HQ expansion
before the integration. 

Assuming that $m_2$ is large and that the meson wave function is localized in the
region $z_1\simeq \Lambda$ we 
obtain the following expression for $\lambda(s_1,s_2,q^2)$ valid at all $q^2$ 
\begin{eqnarray}
\label{lambda}
\lambda(s_1,s_2,q^2)&\to&m_2^4\left({1+\frac{z+m_3}{m_2}}\right)^2
\left[{
\lambda(1,\hat q^2,\hat r^2)
+\frac{2\eta}{m_2}\chi(z)\sqrt{z(z+2m_3)}(1+\hat q^2-\rho^2)
\lambda^{1/2}(1,\hat q^2,\hat r^2)
}\right.\nonumber
\\
&&\left.{+\frac{z(z+2m_3)}{m_2^2}(1+\hat q^2-\hat r^2)^2
+\frac{\eta^2}{m_2^2}z(z+2m_3)\lambda(1,\hat q^2,\hat r^2)}\right]
\end{eqnarray}
where $\hat q^2=q^2/m_b^2$, $\hat r=m_1/m_2$ and $\chi(z)=1-(z+m_3)/2m_2$. 
In the limit $m_2\to\infty$ we can expand the $\lambda(s_1,s_2,q^2)$ in powers
of $1/m_2$. Notice however that an actual expansion parameter is not $1/m_2$ but
rather 
\begin{eqnarray}
\frac{\sqrt{z(z+2m_3)}}{m_2\lambda^{1/2}(1,\hat q^2,\hat r^2)}, 
\end{eqnarray}
and the averaging over the $B$ meson state implies 
$\sqrt{z(z+2m_3)}\simeq \Lambda$. 
Hence the region where the expansion is fastly converging is 
$m_2\lambda^{1/2}(1,\hat q^2,\hat r^2)\gg \Lambda$. This relation can be written as
$|\vec k_1|=\lambda^{1/2}(m_2^2,m_1^2,q^2)/2m_2\gg \Lambda$,
which means that in the rest frame of the $b$ quark the 
daughter quark has a 3-momentum much bigger than $\Lambda$. 

The final expression reads 
\begin{eqnarray}
\label{r2b}
\rho_D(s_1,q^2)\to\lambda^{1/2}(m_1^2,m_2^2,q^2)\frac{m_2}{M_1}\left({1+\frac{z+m_3}{m_2}}\right)
\left[{
1+\frac{z(z+2m_3)}{2m_2^2}\left({1+\frac{8\hat q^2}{3\lambda(1,\hat q^2,\hat r^2)}
}\right)}\right]. 
\end{eqnarray}
The $1/m_Q$ term in the ratio of the bound to free quark distributions is
generated by the $(m_2+z+m_3)/M_1$ term in $\rho_D$.

As we have discussed this linear $1/m_Q$ term contained in 
the box diagram cancels against the $1/m_Q$ terms coming from other terms 
in the expansion (\ref{exp2}). Thus the main contribution of these other 
terms in the series (\ref{exp2}) can be taken into account by performing the 
subtraction in the spectral representation of the box diagram which kills the 
$1/m_Q$ term as follows: 
\begin{eqnarray}
\label{r3}
\rho(s_1,q^2)=\frac{M_1}{\sqrt{s_1}}\rho_D(s_1,q^2).  
\end{eqnarray}
In the HQ limit $z(z+2m_3)=\vec k^2$, so after performing the subtraction we come to the following 
relation  
\begin{eqnarray}
\label{r4}
R(q^2)\equiv \frac{d\Gamma(B\to X_c \ell \bar{\nu}_{\ell}) / dq^2}{d\Gamma(b\to c \ell \bar{\nu}_{\ell}) / dq^2} \to
1+\frac{<\vec k^2>}{2m_2^2}\left({1+\frac{8\hat q^2}{3\lambda(1,\hat q^2,\hat r^2)}
}\right).  
\end{eqnarray}
This expansion is valid in the region of $q^2$ such that $\lambda(1,\hat
q^2,\hat r^2)=O(1)$, i.e. in most 
of the $q^2$ phase space except for the region near zero recoil where $\lambda(1,\hat q^2,\hat r^2) \simeq 0$.  

The expression (\ref{r4}) has the following features:
\begin{itemize}

\item[1.] in the LO the ratio $R(q^2)$ is equal to one and thus the decay rate of the free and the bound quark 
coincide in the HQ limit at all $q^2$. Moreover, the differential distribution also coincide 
within the $1/m_Q$ accuracy in most of the $q^2$ phase space, except for the region near zero recoil. 
This guarantees the absence of the $1/m_Q$ corrections in the ratio of the integrated rates as well. 
Thus, our description is in full agreement with the OPE results within the $1/m_Q$ order; 

\item[2.] since the box diagram represents only a part of the $1/m_Q^2$ corrections, 
we cannot expect the box diagram alone to reproduced correctly the $1/m_Q^2$ term 
in the ratio 
of the integrated rates $\Gamma / \Gamma_0$. 
In fact, the sign of the 
$1/m_Q^2$ correction in eq. (\ref{r4}) turns out to be opposite to the OPE result (cf., e.g., with the results of refs. \cite{bsuv93,mw} at $q^2 = 0$). 
Moreover, the whole 
$1/m_Q^2$ effect in the box diagram is expressed only in terms of $<\vec k^2>$, 
whereas the $1/m_Q^2$ corrections of the whole series contains also $<\hat{V}_1>$ \cite{sim}, 
where $\hat{V}_1$ is the $1/m_Q$ term appearing in the expansion (\ref{Vpot}) of the effective 
potential (e.g., the chromomagnetic operator in QCD).

\end{itemize}

We argue however that it is possible to further modify the spectral 
representation of the box diagram to bring the size of the $1/m_Q^2$ 
term developed by this modified representation in agreement with the OPE result. 
This procedure corresponds to phenomenologically taking into account 
the contribution of other $1/m_Q^2$ terms of the expansion (\ref{exp2}). 

Omitting the anomalous contribution as discussed previously, the differential decay rate reads 
\begin{eqnarray}
\label{uncut}
\frac{d\Gamma}{dq^2}&=&K_0(q^2) 
\int_{(m_1+m_3)^2}^{\infty} ds_1 
\varphi^2(s_1)\frac{s_1-(m_2-m_3)^2}{8\pi^2s_1}\lambda^{1/2}(s_1,m_2^2,m_3^2)
\frac{m_2}{\sqrt{s_1}}
\int_{-1}^{1}\frac{d\eta}{2}\lambda^{1/2}(s_1,s_2,q^2).
\end{eqnarray}
where $s_2$ depends on $\eta$ through eq (\ref{limits}). 

Our goal is to modify the expression (\ref{uncut}) 
in such a way that the LO result and the $1/m_Q$ correction in the integrated rate remain
intact whereas the $1/m_Q^2$ term numerically reproduces the OPE estimate. 
We can allow a strong deformation of the differential $q^2$-distribution at large $q^2$ near
zero recoil, where the HQ expansion is anyway ill-defined. We can also require  
the $1/m_Q^2$ correction in the differential decay rate at $q^2=0$ to exactly reproduce the OPE result. 

Most easily this program may be implemented through the following two steps: first, by 
introducing the factor $F(s_1)=1/(1+\vec k^2/m_Q^2)$ 
which sets the $1/m_Q^2$ term in the {\it differential} rate at $q^2=0$ and, second,
by changing the upper limit in the $s_2$
integration in (\ref{uncut}) to some $s_2^{max}(q^2)$ which tunes the size of the $1/m_Q^2$ 
effects in the {\it integrated} rate:
\begin{eqnarray}
\label{cut}
\frac{d\Gamma}{dq^2}&=&K_0(q^2) 
\int
%\limits_{(m_1+m_3)^2}^{\infty} 
ds_1 
\varphi^2(s_1)\frac{s_1-(m_2-m_3)^2}{8\pi^2s_1}\lambda^{1/2}(s_1,m_2^2,m_3^2)
\frac{m_2}{\sqrt{s_1}}F(s_1)
\int_{-1}^{1} \frac{d\eta}{2} \lambda^{1/2}(s_1,s_2,q^2)\theta\left(s_2<s_2^{max}(q^2)\right). 
\end{eqnarray}
In order not to affect the integrated rate in the LO and the $1/m_Q$ order, 
$s_2^{max}(q^2)$ should satisfy certain properties. 

Assume that the soft wave function $\phi(s_1)$ is localized in the region
$s_1\le s_1^{max} \simeq (m_2+m_3+\gamma)^2$ where $\gamma$ is a constant of order $\Lambda$ which does not scale with $m_Q$. 
Let us determine $q_0^2$ through the equation 
\begin{eqnarray}
s_2^+(s_1^{max},q_0^2)=s_2^{max}(q_0^2)
\end{eqnarray}
where $s_2^+$ is the maximal value of $s_2$ corresponding to $\eta=1$ in (\ref{limits}). Furthermore, assume that $s_2^{max}(q^2)$ decreases with $q^2$, and take into account that $s_2^+(s_1,q^2)$ is a monotonous rising function of both $s_1$ and $q^2$. 
Then, at $q^2<q_0^2$ for all $s_1<s_1^{max}$ one finds the relation 
$s_2^+(s_1,q^2)<s_2^{max}(q^2)$, and thus
the $q^2$-distribution does not feel the presence of the cut at all. 
For $q^2>q_0^2$ the cut becomes really effective and strongly
influences the $q^2$-distribution. In order these changes in the cut $q^2$-differential 
distribution not to change the integrated rate in the LO and $1/m_Q$ order, we need the 
$q_0^2$ to be not far from zero recoil such that the corresponding $\omega_0=1+O(1/m_Q)$. 
Choosing the cut in the form 
\begin{eqnarray}
\label{rcut}
\sqrt{s_2^{max}(q^2)}=m_1+m_3+a\left(m_2-m_1-\epsilon-\sqrt{q^2}\right), 
\end{eqnarray}
where $\epsilon\simeq \Lambda$ and $a$ is a rising function of $m_Q$, satisfies 
these requirements\footnote{One could choose a more 
sophisticated parameterization of $s_2^{max}(q^2)$ 
to reproduce a correct $q^2$-behaviour of $d\Gamma/dq^2$ near zero recoil point. For instance, 
taking into account that the lightest final meson is pseudoscalar, we can write  
$\sqrt{s_2^{max}(q^2)}=m_1+m_3+a_P(m_Q) (M_{P_Q}-M_{P_{Q'}}-\sqrt{q^2})^{3/2}$ yielding 
the correct behavior near $\sqrt{q^2}=M_{P_Q}-M_{P_{Q'}}$ where only one $P-$wave decay channel  
$P_Q\to P_{Q'} \ell \bar{\nu}_{\ell}$ is opened. In addition, in the heavy quark limit the $S-$wave transition 
$P_Q\to V_{Q'} \ell \bar{\nu}_{\ell}$ requiring another functional dependence 
$\sqrt{s_2^{max}(q^2)}=m_1+m_3+a_V(m_Q)(M_{P_Q}-M_{P_{Q'}}-\sqrt{q^2})^{1/2}$
is opened at $\sqrt{q^2}=M_{P_Q}-M_{V_{Q'}}$ with only small delay in $q^2$ of order $\Lambda^2$. 
So the effects of opening this channel are even more important and should be also taken into account. 
However in the region of large $q^2$ with few opened channels the inclusive consideration 
is anyway not working properly, and taking into account such subtle effects is beyond the 
accuracy of the method. So in numerical calculations we proceed with the phenomenological 
cut provided by eq (\ref{rcut}).}. 
 
The parameter $\epsilon$ accounts for a mismatch between the quark and the hadron threshold, and 
the form of  
$a(m_Q)$ can be found from fitting the size of the $1/m_Q^2$ corrections in the 
integrated rate to the OPE prediction.    
Notice also that the $q^2$ distributions obtained through the cut expression are even 
more realistic than those obtained from the uncut spectral representations. 
Numerical values used for the description of the distributions are given in the following section. 

Mostly important for us however is that these improvements on the $q^2$-differential
distributions by approximate account of higher order graphs 
affect only moderately the $M_X$-distribution in $B\to X_c \ell \bar{\nu}_{\ell}$ 
(as well as the photon lineshape 
in the rare $B\to X_s\gamma$ decay), which are thus mostly determined by the $B$-meson wave
function. The latter controlls long-distance effects also in exclusive transitions. 

This property allows us to obtain a realistic energy distribution and other observables 
through the soft wave function of the heavy meson. Thus we do not need to introduce 
any unknown 'smearing function' describing the motion of the $b$ quark inside the $B$ 
meson, but rather directly calculate the effects of the $b$ quark motion with the 
soft meson wave function.  

\section{Differential distributions}

In previous sections we have considered in detail the integrated rate and the differential 
distribution $d\Gamma/dq^2$. We expect the obtained spectral representations of these quantities 
to describe some essential features of the exact quantities. The $1/m_Q$ expension of these rates 
reproduce the known OPE results allowing to express nonperturbative parameters through the B-meson 
soft wave function. Interesting results can be obtained also for other  
differential distributions, such as the $M_X$-distribution and the lepton energy spectrum  
in SL decays. 

For instance, neglecting the 
radiative corrections in the free-quark decay, 
which is the LO process within the OPE framework, 
one finds the $M_X$ distribution in the form $\delta(M_X-m_c)$. 
Inclusion of the $1/m_Q$ corrections yields a singular series 
containing derivatives of the $\delta$ function. For the interpretation of these
results one needs an introduction of a smearing function. In the quark model 
the $M_X$-spectrum is smeared because of the Fermi-motion of the $b$ quark in the $B$
meson and the shape of the $M_X$-spectrum is calculable through the $B$ meson wave function.  

The meson wave function $\varphi(s)$ can be written as follows \cite{m} 
\begin{equation}
\label{vertex}
\varphi(s) = \frac{\pi}{\sqrt{2}} \frac{\sqrt{s^2 - (m_Q^2 -
m_{\bar{q}}^2)^2}} {\sqrt{s - (m_Q - m_{\bar{q}})^2}} \frac{w(k^2)}{s^{3/4}}, 
\end{equation}
where $k = |\vec k|=\lambda^{1/2}(s, m_Q^2, m_{\bar{q}}^2) / 2 \sqrt{s}$ and
$w(k^2)$ is the ground-state $S$-wave radial wave function, normalized as
$\int_0^{\infty} dk k^2 |w(k^2)|^2 = 1$. 

As shown by the analysis of the exclusive form factors within the dispersion approach \cite{bm}, for obtaining a good description of the lattice results on the $B\to \pi,\rho$ form factors at 
large $q^2$ it is sufficient to take into account properly only the confinement scale effects, i.e. to 
assume a simple exponential parameterization of the radial wave function in the form 
$w_B(\vec k^2)\simeq \exp(-\vec k^2/2\beta_B^2)$. 
Then, the quark masses $m_u=0.23\; GeV$ and $m_b=4.85\;GeV$ as well as the value $\beta_B=0.54 \;GeV$
have been determined from fitting the lattice data 
on the form factors. This value of $\beta_B$ corresponds to $\lambda_1 = - <\vec{k}^2> \simeq -0.44 ~ GeV^2$.  
Finally, the mass of the c-quark is taken equal to $m_c = 1.35 ~ GeV$ in agreement with current estimates of the difference $m_b - m_c$ ($\simeq 3.5 ~ GeV$). 

\begin{figure}[htb]
\vspace{0.25cm}
\begin{center}
\epsfig{file=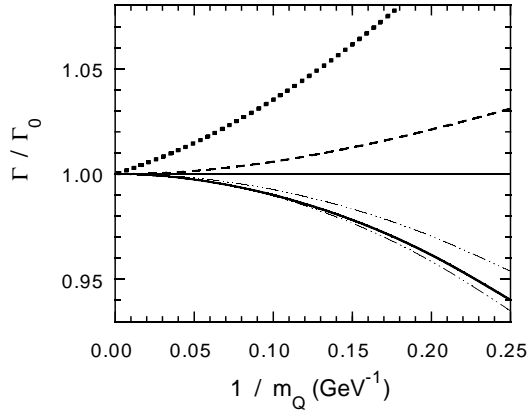,width=7cm}
\vspace{0.25cm}
\caption{ The ratio of the integrated rates of the bound-to-free quark SL 
decay $Q \to Q' \ell \bar{\nu}_{\ell}$ vs the inverse HQ mass  
$1/m_Q$ at fixed value of $m_{Q'} / m_Q = m_c / m_b = 0.28$. Dotted line - the rate 
calculated from the initial spectral representation of the box diagram which 
contains $1/m_Q$ correction, dashed - with the proper subtraction killing 
the $1/m_Q$ term but without tuning the size of the $1/m_Q^2$ effects. Solid 
- final result which also includes the cut bringing the size of the 
$1/m_Q^2$ effects in agreement with the OPE prediction.  Upper and lower 
dot-dashed lines are the OPE results (\ref{OPE_total}) corresponding to $\lambda_1 = 0$ and $\lambda_1 = -0.6 ~ GeV^2$ (with $\lambda_2 = 0.12 ~ GeV^2$), respectively.
\label{fig:rate}}
\end{center}
\end{figure}

The parameters of the cut have been chosen according to the criteria of the previous section, obtaining 
$\epsilon=(m_Q-m_{Q'})-(M_Q-M_{Q'})$ and $a = 1.82 + 0.029 m_Q$, where $m_{Q'}$ is the mass of the parent heavy quark and 
$M_{Q'}$ is the final meson lowest
mass. Effectively, this means that $q^2_{max}=(M_Q-M_{Q'})^2$.  
The corresponding integrated decay rate $\Gamma$ 
is plotted in Fig. \ref{fig:rate} as a function of $1/m_Q$ and compared with 
the OPE predictions (\ref{OPE_total})
\footnote{To compare the calcuated integrated rate $\Gamma$ as a 
function of $m_Q$ with the OPE result 
within the $1/m_Q^2$ order, it is reasonable to adopt the expansions of 
the hadron masses $M_Q$ and $M_{Q'}$ also up to the second order in 
$1/m_Q$. Then one finds 
$\epsilon = -\frac{1}{2}(\lambda_1 + 3 \lambda_2) \cdot (1/m_{Q'} - 1/m_{Q})$. 
Setting $\lambda_1 = -0.44 ~ GeV^2$ and $\lambda_2 = 0.12 ~ GeV^2$, 
yields $\epsilon = 0.10 / m_Q$ at $m_{Q'} / m_Q = m_c / m_b \simeq 0.28$.
We point out that in calculations for the real $B$ decays we use experimental values of 
hadronic masses (involving all orders in $1/m_Q$.)}. 
One can clearly see that the calculations based on both the unsubtracted $\rho_D$ (\ref{r2}) and subtracted $\rho$ (\ref{r3}) 
densities predict a larger rate for a bound heavy quark that contradicts OPE, 
whereas the introduction of the 
phenomenological cut $s_2^{max}(q^2)$ (\ref{rcut}) brings our QM predictions in perfect 
agreement with the standard OPE in the whole range of considered values of $m_Q$.  

\begin{figure}[htb]
\vspace{0.25cm}
\begin{center}
\epsfig{file=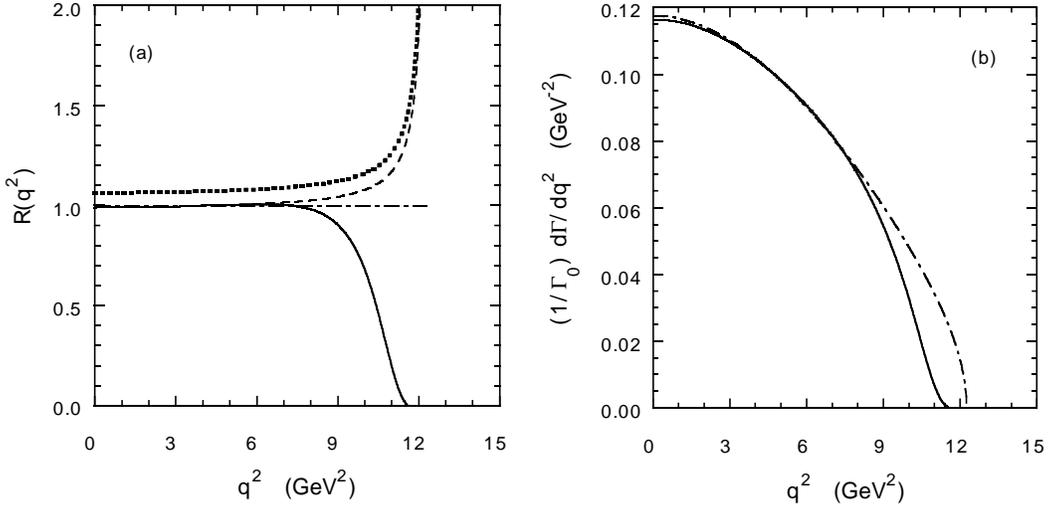,width=14cm}
\vspace{0.25cm}
\caption{Distribution $d\Gamma/dq^2$ in $B \to X_c \ell \bar{\nu}_{\ell}$ decays vs the 
squared four-momentum transfer $q^2$: (a) ratio of the bound-to-free quark decay 
$R(q^2) \equiv (d\Gamma / dq^2) / (d\Gamma_0 / dq^2)$, 
notation of lines same as in Fig. \protect\ref{fig:rate}. (b) Differential distribution 
in the bound 
(solid) and free (dot-dashed) SL quark decay. Parameters of the cut (\ref{rcut}) are 
$a=1.96$ and $\epsilon=0.091\;GeV$. 
\label{fig:rq2}}
\end{center}
\end{figure}

Figure \ref{fig:rq2} shows the influence of the cut upon the 
$q^2$-distribution for the $B \to X_c \ell \bar{\nu}_{\ell}$ decay. As already discussed, the introduction of the $q^2$-dependent cut in the spectral 
representation does not change the differential $q^2$ distributions at small $q^2$ but it strongly 
affects the region of large $q^2$. In particular, the cut provides the vanishing of 
$d\Gamma/dq^2$ at the {\it physical} threshold $q^2=(M_B-M_D)^2=11.6\;GeV^2$. Let us point out again that 
this cutting procedure does not affect the {\it integrated} rate at the leading and subleading 
$1/m_Q$ orders. Note also that the differential distributions $d\Gamma/dq^2$ given by the dashed and solid lines in Fig \ref{fig:rq2}(a), 
are equal to each other at $q^2=0$ and match the OPE result for $d\Gamma/dq^2(q^2=0)$.
 
\begin{figure}[htb]
\vspace{0.25cm}
\begin{center}
\epsfig{file=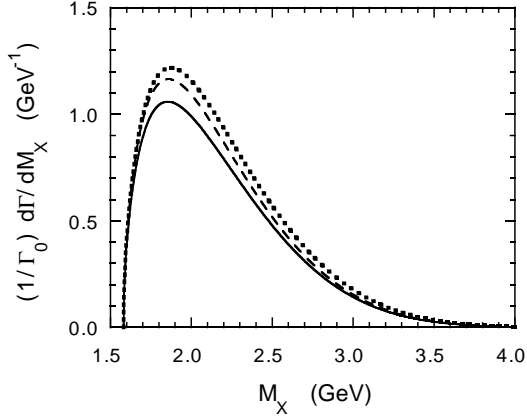,width=7cm}
\vspace{0.25cm}
\caption{Distribution $d\Gamma / dM_X$ in $B \to X_c \ell \bar{\nu}_{\ell}$ decays vs the invariant mass of the produced hadronic system $M_X$. Lines same as in Fig. \protect\ref{fig:rate}.  
\label{fig:rmx}}
\end{center}
\end{figure}

Mostly interesting seems to be the calculated $M_X$-distribution, reported in 
Fig.\ref{fig:rmx}. Our result should be compared with the LO OPE result $\delta(m_X-m_c)$. 
One can see that already the box diagram of the quark model provides a smooth and reasonable 
(beyond the resonance region) distribution, which is only moderately affected by a proper 
account of the subleading $1/m_Q$ effects. At large $M_X$ the calculated distribution does 
not require any additional smearing and can be directly compared with the experimental results.

The double differential distribution $d^2\Gamma/dEdq^2$ is given by the following expression 
\begin{eqnarray}
\label{double}
\frac{d^2\Gamma}{dEdq^2}&=& \frac{G_F^2|V_{21}|^2}{128\pi^3}
\int
%\limits_{(m_2+m_3)^2}^{\infty}
ds_1\varphi^2(s_1)
\frac{s_1-(m_2-m_3)^2}{8\pi^2s_1}\lambda^{1/2}(s_1,m_2^2,m_3^2)\frac{F(s_1)}{m_2^2}
\int_{-1}^{1}\frac{d\eta}{2} \theta\left(q_0>E+\frac{q^2}{4E}\right)
\theta\left(s_2<s_2^{max}(q^2)\right)\nonumber\\
&&\times
\left(2q^2 w_1(s_1,s_2,q^2)
+[4E(q^0-E)-q^2] w_2(s_1,s_2,q^2)+2q^2(2E-q^0) w_3(s_1,s_2,q^2) \right). 
\end{eqnarray}
In this formula $q^0$ is expressed in terms  of the integration variables,  namely: $q^0=(s_1+q^2-s_2)/{2\sqrt{s_1}}$. 
The functions $w_i(s_1,s_2,q^2)$ have the form (for more details see Appendix)
\begin{eqnarray}
q^2 w_1+\frac{\vec q^2}{3} w_2=\frac{4}{3}C(m^2_1,m^2_2,q^2),\qquad 
w_1=4(m_1^2+m_2^2-q^2)-16\beta,\qquad 
w_3=8\sqrt{s_1}(1-\alpha_1-\alpha_2),
\end{eqnarray}
with $\beta$, $\alpha_1$, $\alpha_2$ given by eqs (36-38)  
in \cite{m} and  
$C(m^2_1,m^2_2,q^2)$ defined in (\ref{c}). 

\begin{figure}[htb]
\vspace{0.25cm}
\begin{center}
\epsfig{file=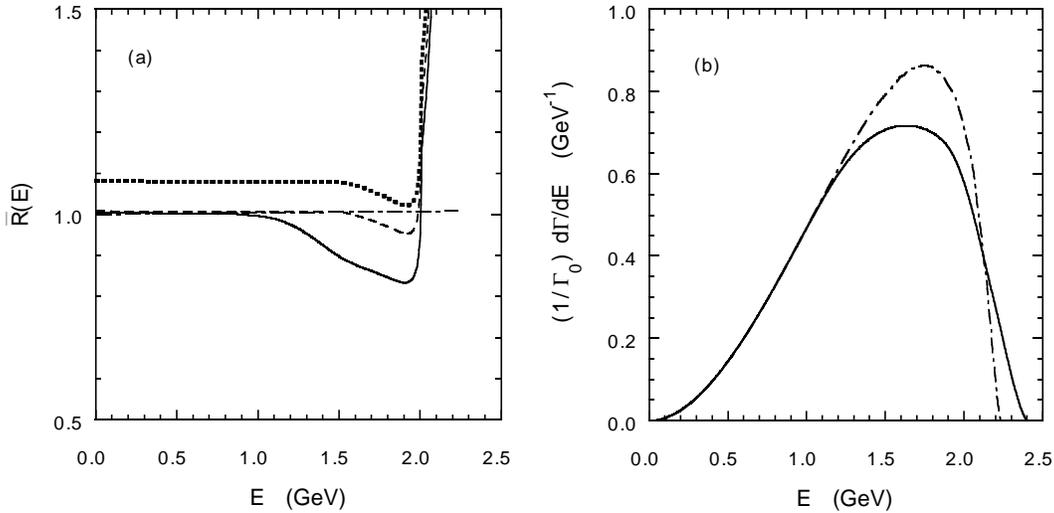,width=14cm}
\vspace{0.25cm}
\caption{Distribution $d\Gamma / dE$ in $B \to X_c \ell \bar{\nu}_{\ell}$ decays vs the lepton energy $E$:
(a) ratio of the bound-to-free quark decay $\bar{R}(E) \equiv (d\Gamma/dE)/(d\Gamma_0 / dE)$, lines same as in Fig. \protect\ref{fig:rate}, (b) Our QM result (solid) vs free-quark result (dot-dashed).
\label{fig:re}}
\end{center}
\end{figure}

The electron spectrum $d\Gamma/dE$ is obtained by integrating (\ref{double}) over $q^2$. 
Fig. \ref{fig:re} plots the results of the calculations. 
Here the effects of the subleading $1/m_Q$ orders are more pronounced but 
nevertheless lead only to a moderate change of the quark model box-diagram 
result. We also compare the quark model prediction with the electron spectrum 
in the free-quark decay process.

\section{Conclusion}

We have proposed a new approach to the description of the quark-binding effects in the 
inclusive decays of heavy mesons. Our approach is based on the dispersion formulation of the 
relativistic quark model and allows one to express kinematical distributions in inclusive 
decays of heavy mesons in terms of the heavy meson soft wave function. 
This soft wave function describes long-distance effects both in exclusive and incluisve 
processes. 

Our main results are as follows:
\begin{itemize}

\item[1.] we have analysed the hadronic tensor in the quark model. We argue that 
the diagrammatic representation of the hadronic tensor in the quark model yields
an expansion in inverse powers of the heavy quark mass as well as the standard OPE. However 
in distinction to the standard OPE which is based on the free quark decay as the LO process, the 
LO process in the quark model is described by the box diagram 
with a free $Q\bar q$ in the final state. This yields some specific features of the hadronic 
tensor calculated within the QM, both negative and positive.
On the one hand, a consideration based on the box diagram alone reproduces the correct 
LO result but contains also $1/m_Q$ corrections. These $1/m_Q$ terms are cancelled 
against contributions of other graphs thus leading to the agreement with the standard 
OPE result. Hence the effects of the subleading diagrams should be taken into account 
for obtaining a consistent approach.
On the other hand, the hadronic tensor calculated from the box diagram is a regular function in 
the kinematically allowed region of all variables, as well as all the subleading order $1/m_Q$ terms. 
This feature makes the quark model calculation of the hadronic tensor 
very suitable for describing the differential distributions; 

\item[2.] we have constructed the double spectral representation of the hadronic tensor for the $B \to X_c \ell \bar{\nu}_{\ell}$ decay
in terms of the $B$ meson soft wave function and the double spectral density of the box diagram, 
and analysed its $1/m_Q$ expansion in the case of the heavy-to-heavy inclusive transition. 
The spectral representation is further modified in order to take into account essential 
effects of the other diagrams contributing in subleading orders. Namely,  
the subtraction term in this dispersion representation is determined such that the $1/m_Q$ correction 
in the integrated SL rate is absent in agreement with the OPE result. Furthermore, a phenomenological 
cut is introduced in the spectral representation to bring the size of the $1/m_Q^2$ terms in the 
differential $q^2$-distribution at $q^2=0$ and in the integrated rate into full agreement with the 
OPE. Thus our representation of the hadronic tensor obeys the OPE predictions in the regions 
where the latter are expected to be valid; 

\item[3.] we have obtained numerical results on differential distributions in inclusive $B \to X_c \ell \bar{\nu}_{\ell}$ decays using the 
$B$-meson wave function and other quark model parameters previously determined from the 
description of exclusive meson transition form factors within the dispersion approach. So, basically our predictions are parameter-free.
We notice that 
modifications of the spectral representations which take into account the subleading $1/m_Q$ 
effects 
within the box-diagram representation, introduce some uncertainties in our results. However, they  
do not affect our predictions strongly, and the main features of the 
inclusive distributions are determined by the soft meson wave function. Moreover, the size of 
the 
subleading corrections is in perfect agreement with the OPE result for the integrated rate, 
and we expect 
to describe also these subleading effects in differential distributions in a proper 
quantitative way.

\end{itemize} 

The proposed approach can be applied to the inclusive $B\to X_{u,s}$ transitions.  
In particular it is especially suitable for the description of the photon line shape in $B \to X_s \gamma$ decays.  
However, certain subtleties in heavy-to-light transitions 
compared with the heavy-to-heavy decays emerge. They  
are mostly connected with the fact that in heavy-to-light transition the kinematically allowed 
$q^2$-interval of the hadron SL decay is larger than that of the quark decay, while in case of the 
heavy-to-heavy transitions the situation is just opposite and the $q^2$-region of the quark decay is larger. 
This feature requires a detailed analysis of the $q^2$-region near zero recoil in heavy-to-light 
inclusive decays. 

It is also worth noting that our approach takes into account only 
non-perturbative effects in inclusive decays of heavy mesons. 
Perturbative corrections have been ignored. So, for 
comparing our results with the experimental differential distributions perturbative corrections 
should be also included into consideration. 

We are going to address these issues in a separate work. 

\acknowledgements
The authors are grateful to V. Anisovich and B. Stech for stimulating discussions.

\newpage

\section{Appendix: Calculation of the double differential distribution}  

Here some technical details of calculaitng the $d^2\Gamma/dq^2dE$ are provided. 
We start from the expression (\ref{ad}) which gives the double discontinuity 
of the box diagram. 
The trace corresponding to the $V-A$ current has the form  
\begin{eqnarray}
\label{a1}
{\rm Sp}\left(\gamma_5\,(m_3-\hat k_3)
\gamma_5\,
(m_2+\hat k_2)\gamma_\mu(1-\gamma_5)(m_1+\hat k_1)\gamma_\nu(1-\gamma_5)
(m_2+\hat k_2)\right)=2\left(s_1-(m_2-m_3)^2\right)\cdot \bar w_{\mu\nu},    
\end{eqnarray} 
where $\bar w_{\mu\nu}$ is the trace over the free-quark loop 
\begin{eqnarray}
\label{a2}
\bar w_{\mu\nu}={\rm Sp}\left(
(m_2+\hat k_2)\gamma_\mu(1-\gamma_5)(m_1+\hat k_1)\gamma_\nu(1-\gamma_5)
\right)=8[k_{1\mu}k_{2\nu}+k_{2\mu}k_{1\nu}-g_{\mu\nu}k_1k_2
+i\epsilon_{\mu\nu\alpha\beta}k_{2\alpha}q_\beta].     
\end{eqnarray} 

One finds the following useful relation 
\begin{eqnarray}
\label{a3}
-\frac{1}{3}\left(q^2g_{\mu\nu}-q_\mu q_\nu\right)\bar w_{\mu\nu}=
\frac{8}{3}(k_1k_2\;q^2+2 k_1q\;k_2q)
=\frac{4}{3}\left[(m_2^2-m_1^2)^2+q^2(m_1^2+m_2^2)-2q^4\right]
=\frac{4}{3}C(m^2_1,m^2_2,q^2). 
\end{eqnarray}
The integration over $dk_1dk_2\dots$ in eq (\ref{ad}) yields 
\begin{eqnarray}
\label{a4}
\left(\tilde A_D \right)_{\mu\nu}=
\frac{\pi\theta(\dots)\left(s_1-(m_2-m_3)^2\right)}
{\lambda^{1/2}(m_2^2,m_1^2,q^2)}w_{\mu\nu}(\tilde p_1,\tilde q)
\end{eqnarray}
where $w_{\mu\nu}$ is represented in terms of the 'dispersion momenta'  
$\tilde p_1$ and $\tilde q=\tilde p_1-\tilde p_2$ 
($\tilde p_1^2=s_1$, $\tilde p_2^2=s_2$, $\tilde q^2=q^2$) as follows 
\begin{eqnarray}
\label{a5}
w_{\mu\nu}(\tilde p_1,\tilde q)=-g_{\mu\nu}w_1
+\frac{\tilde p_{1\mu}\tilde p_{1\nu}}{s_1}w_2
+i\epsilon_{\mu\nu\alpha\beta}\frac{\tilde p_{1\mu}}{\sqrt{s_1}}\tilde q_\beta w_3
+\frac{\tilde p_{1\mu}\tilde q_\nu+\tilde p_{1\nu}\tilde q_\mu}{\sqrt{s_1}}w_4
+\tilde q_\mu \tilde q_\nu w_5. 
\end{eqnarray}
Notice that 
\begin{eqnarray}
\label{a6}
-\frac{1}{3}\left(q^2g_{\mu\nu}-q_\mu q_\nu\right)w_{\mu\nu}=
q^2w_1+\frac{1}{3}\vec q^2 w_2, 
\end{eqnarray}
where 
\begin{eqnarray}
\label{a61}
|\vec q|=\frac{\lambda^{1/2}(s_1,s_2,q^2)}{2\sqrt{s_1}}, \qquad
q^0=\frac{s_1+q^2-s_2}{2\sqrt{s_1}}.
\end{eqnarray}

Comparing (\ref{a3}) and (\ref{a6}) one finds 
\begin{eqnarray}
\label{a7}
q^2 w_1+\frac{1}{3}\vec q^2\ w_2=\frac{4}{3}C(m^2_1,m^2_2,q^2).
\end{eqnarray}

As the next step, performing the convolution with the leptonic tensor gives 
the double differential distribution 
\begin{eqnarray}
\label{a8}
\frac{d^2\Gamma}{dEdq^2}&=& \frac{G_F^2|V_{21}|^2}{128\pi^3}
\int
%\limits_{(m_2+m_3)^2}^{\infty}
ds_1\varphi^2(s_1)
\frac{s_1-(m_2-m_3)^2}{8\pi^2s_1}\lambda^{1/2}(s_1,m_2^2,m_3^2)\frac{F(s_1)}{m_2^2}
\int_{-1}^{1}\frac{d\eta}{2} \theta\left(q_0>E+\frac{q^2}{4E}\right)
\theta\left(s_2<s_2^{max}(q^2)\right)\nonumber\\
&&\times
\left(2q^2 w_1(s_1,s_2,q^2)
+[4E(q^0-E)-q^2] w_2(s_1,s_2,q^2)+2q^2(2E-q^0) w_3(s_1,s_2,q^2) \right). 
\end{eqnarray}
In this formula $s_2$ is connected with $\eta$ through eq (\ref{limits}), 
and $q_0$ is given in terms of the integration variables by eq (\ref{a61}).

The eq (\ref{a8}) is based on the box diagram and also includes modifications 
which tune the size of the $1/m_Q^2$ corrections as explained in the text.  

By virtue of the relations  
\begin{eqnarray}
&&\int dE\theta(q^0>E+q^2/4E)=|\vec q|,\\
&&\int E dE\theta(q^0>E+q^2/4E)=q^0|\vec q|/2,\\
&&\int E^2 dE\theta(q^0>E+q^2/4E)=\frac{1}{12}|\vec q|(3q_0^2+\vec q^2).
\end{eqnarray}
one finds 
\begin{eqnarray}
\int dE 
\{2q^2 w_1+[4E(q^0-E)-q^2]w_2+2q^2(2E-q^0)w_3 \}\theta(q^0>E+q^2/4E)=
2|\vec q|(q^2w_1+\frac{1}{3}\vec q^2w_2). 
\end{eqnarray}
Thus, integrating the double differential distribution (\ref{a8}) over $E$ 
and using (\ref{a7}) gives $d\Gamma/dq^2$ in the form (\ref{cut})
\begin{eqnarray}
\label{a10}
\frac{d\Gamma}{dq^2}&=& \frac{G_F^2|V_{21}|^2}{96\pi^3m_2^3}C(m_2^2,m_1^2,q^2)
\int
\limits_{(m_2+m_3)^2}^{\infty}
ds_1\varphi^2(s_1)
\frac{s_1-(m_2-m_3)^2}{8\pi^2s_1}\lambda^{1/2}(s_1,m_2^2,m_3^2)\nonumber\\
&&\times
F(s_1)\frac{m_2}{\sqrt{s_1}}
\int_{-1}^{1}\frac{d\eta}{2}
\lambda^{1/2}(s_1,s_2,q^2)\theta\left(s_2<s_2^{max}(q^2)\right). 
\end{eqnarray}
It is clear that for the calculation of $d\Gamma/dq^2$ 
we do not need to know all $w_i$, but only their linear combination 
(\ref{a7}). 

On the other hand, for calculating the electron spectrum 
$d\Gamma/dE$ we need all the functions $w_i$. The latter read 
\begin{eqnarray}
q^2 w_1+\frac{\vec q^2}{3} w_2&=&\frac{4}{3}C(m^2_1,m^2_2,q^2),\nonumber\\
 w_1&=&8k_1k_2-16\beta=4(m_1^2+m_2^2-q^2)-16\beta,\nonumber\\
 w_3&=&8\sqrt{s_1}(1-\alpha_1-\alpha_2), 
\end{eqnarray}
where $\beta$, $\alpha_1$, $\alpha_2$ are given by eqs (36-38) 
of \cite{m}. 

It might be interesting to note that at $q^2<0$, and using the reference frame 
$q_+=0$, $p_{1\perp}=0$ ($q^2=-q_\perp^2$) one obtains 
\begin{eqnarray}
\beta=-\left(k_\perp^2-\frac{(k_\perp q_\perp)^2}{q_\perp^2}\right),\qquad
1-\alpha_1-\alpha_2=1-x_3, 
\end{eqnarray}
where $x_3$ and $k_\perp$ are the $(+)$ and $(\perp)$ components of the spectator
quark momentum, respectively. In the heavy meson one finds 
\begin{eqnarray}
\label{a11}
\beta\simeq \Lambda^2,\qquad x_3\simeq \Lambda/m_Q.
\end{eqnarray}
Although at $q^2>0$ the 
interpretation of $\beta$ and $\alpha_1+\alpha_2$ in terms of 
$k_\perp$ and $x_3$ is not straightforward, the estimates 
(\ref{a11}) remain valid also at $q^2>0$. This means that in $B \to X_c \ell \bar{\nu}_{\ell}$ decays our $w_i$'s differ only slightly from the corresponding free-quark expressions.
\end{document}